\documentclass[12pt]{article}
\usepackage[utf8]{inputenc}
\usepackage[english]{babel} 
\usepackage{color}
\usepackage{amssymb}
\usepackage{amsmath}
\usepackage{amsfonts}
\usepackage[font={footnotesize}]{caption}
\usepackage{mathptmx}
\usepackage{hyperref}
\usepackage{graphicx}
\usepackage{wrapfig}
\usepackage{float}
\usepackage{multicol}
\usepackage{anysize}
\marginsize{1.5cm}{1.5cm}{2.5cm}{3.5cm}
\date{}
\begin{document}

\title{\textbf{Logistic map trajectory distributions: Renormalization-group, entropy and criticality at the transition to chaos}}


\author{A. Diaz-Ruelas\thanks{Max Planck Institute for the Physics of Complex Systems, Nöthnitzer Str. 38, D-01187 Dresden, Germany \ (diazruelas@pks.mpg.de)}, F. Baldovin \thanks{INFN-Dipartimento di Fisica,  Università di Padova, Via Marzolo 8, I-35131 Padova, Italy, \ (baldovin@pd.infn.it)}, A. Robledo\thanks{Instituto de Física, Universidad Nacional Autónoma de México, Apartado Postal 20-364, México 01000 D.F., Mexico, \ (robledo@fisica.unam.mx)}}



\maketitle


\begin{abstract}
We study the evolution of the probability density of ensembles of iterates of the logistic map that advance towards and finally remain at attractors of representative dynamical regimes. We consider the mirror families of superstable attractors along the period-doubling cascade, and of chaotic-band attractors along the inverse band-splitting cascade. We examine also their common aperiodic accumulation point. The iteration time progress of the densities of trajectories is determined via the action of the Frobenius-Perron (FP) operator. As a difference with the study of individual orbits, the analysis of ensembles of positions offers a viewpoint from which the nonlinear dynamical features of this iconic model can be better characterized in statistical-mechanical terms. The scaling of the densities along the considered families of attractors conforms to a renormalization-group (RG) structure, while their entropies are seen to attain extrema at the fixed points of the RG flows. Additionally, this entropy as a function of the map control parameter displays the characteristic features of an equation of state of a thermal system undergoing a second-order phase transition. We discuss our results.
\end{abstract}


\maketitle

\begin{quotation}
Already few decades ago it had been a common general commentary within the Complex Systems community that observations of complex systems in nature appear to indicate, in the language of nonlinear dynamics, that their conduct is as if they evolve at the `edge of chaos'. Likewise, the same community nowadays often shares the general commentary that the observations of complex systems in nature seem to imply, in the language of statistical mechanics, that they thrive in a state of `criticality'. Interestingly, as we describe here, these two paradigms appear to be equivalent at the transition to chaos displayed by the archetypal nonlinear dynamical model, the quadratic map. To see this we consider two families of attractors, the supercycles along the period-doubling cascade and the band-splitting (Misiurewicz) points along the chaotic-band cascade, together with their joint accumulation point at the transition to and out of chaos. With their invariant densities in hand (provided by the Frobenius-Perron method) a familiar Renormalization Group picture appears, while the uncomplicated task of evaluating their entropies is an opportunity to be taken. First of all, the fixed points are identified as entropy extrema. For period one the entropy vanishes reaching its minimum possible value, while the entropy for the single chaotic band attains the maximum value. The entropy for the nontrivial fixed point at the transition to chaos is maximum for all supercycles and minimum for all Misiurewicz points. Secondly, the entropy of the invariant densities grows monotonically from period one through all supercycles, and all Misiurewcz points to the final single-band chaotic attractor. But most remarkably, when the collection of entropies for the two families of attractors is viewed along the values of control parameter of the map the familiar pattern appears of a statistical-mechanical two-phase system separated by a continuous phase transition, an equation of state containing a critical point.
\end{quotation}


\section{\label{Sec:Introduction_FPFP}Introduction}

The conventional approach to study the dynamics of low-dimensional nonlinear systems, e.g., iterated maps of the interval, is to look at the asymptotic properties of single orbits \cite{Schuster1, Hilborn1, Beck1}. In contrast here we analyze both the transient and the asymptotic behavior of the probability distribution, or density, associated with ensembles of orbits. This change of perspective is comparable to that present in the description of classical normal diffusion, in which tracking the dynamics of single particles via the Langevin equation \cite{Chaikin1} is reformulated into a partial differential equation for the evolution of the probability density of finding a particle at a specific position and time, i.e., the Fokker-Planck equation \cite{Risken1}. This substitution in object of study helps to illuminate the inner workings behind statistical-mechanical descriptions \cite{Cvitanovic1}. Motivated by potential statistical-mechanical insight gain, and as a difference with normal diffusion, here we look at the evolution of the densities of trajectories that take place in situations governed by nonlinear dynamics. We choose to examine the familiar period-doubling route to chaos in dissipative systems together with its companion sequence of chaotic-band splitting attractors \cite{Schuster1, Hilborn1, Beck1}. For practical reasons a convenient setting for our planned investigation is the standard logistic map. Unless specifically stated, we describe the evolution of densities of a uniformly distributed set of initial conditions along the interval of definition of the map. As the trajectories evolve towards the attractors the densities advance likewise to a final stage that reflects the distinct visiting order of attractor positions or bands in unimodal maps \cite{Schuster1, Hilborn1, Beck1}. As anticipated the known self-affine properties displayed by these families of attractors \cite{Schuster1, Hilborn1, Beck1} manifest also in their densities and we take advantage of this to formulate an appropriate Renormalization Group (RG) transformation for which the density at the RG nontrivial fixed point corresponds to the transition into or out of chaos. The densities  for the trivial fixed points are those for period one and single band attractors. But also the entropies associated with the densities can be readily determined and these are found to be extrema for the RG fixed points. The overall picture obtained is that of a statistical-mechanical system in the vicinity of a critical point.
\\

\textit{Background recall.} The density of trajectories $\rho_t(x)$ at positions $x\in I$ and iteration time $t$ under the action of a given map $f(x)$ defined for the phase-space interval $I$ can be constructed directly from an initial density by means of a linear operator approach, This operator, known as the transfer or Frobenius-Perron operator \cite{Frobenius-Perron1}, acts on arbitrary densities and drives them forward in time. It is defined by the action

\begin{equation}
\rho_{t}(x) = \mathcal{L}^{(t)}\rho_0(x),
\label{Eq:Perron-FrobeniusOperator_Def}
\end{equation}
which is written in explicit form as

\begin{equation}
\rho_t(x) = \int_{I} \delta[y-f^{(t)}(x_0)] \rho_0(y) dy, \label{Eq:FPequation}
\end{equation}
where $x=f^{(t)}(x_0)$ is $f(x_0)$ composed $t$ times with itself. Eq. (\ref{Eq:FPequation}) is called the Frobenius-Perron equation and $\delta[y-f^{(t)}(x_0)]$ is the singular kernel of its associated linear operator \cite{Cvitanovic1}. \noindent Even though the evolution of trajectories under $f(x)$ is nonlinear there is a linear relation between densities via the integral operation. Together with Eq. (\ref{Eq:FPequation}) we require

\begin{equation}
\int_{f^{(t)}(I)} \rho_t(x) dx = \int_I \rho_0(x) dx, \nonumber
\end{equation}
for arbitrary densities $\rho_0(x)$ of initial conditions distributed over $I$. This is equivalent to the conservation of the total Lebesque measure of $I$ under $f(x)$, or, in other words, the initial number of trajectories is preserved. We consider the relation above to be valid for the dissipative case, as employed here. \cite{Cvitanovic1,Lasota1}.\\ 

In the following Section 2 we particularize the Frobenius-Perron approach to the logistic map, and in the next Section 3 we present the resultant densities of ensembles of trajectories that first proceed towards and then evolve within the superstable attractors, the band-splitting or Misiurewicz points, and their common accumulation point, the Feigebaum attractor \cite{Schuster1, Hilborn1, Beck1}. We indicate there the development of a larger (than consecutive iteration $t$) time scale, of the form $\tau=N2^n, N=1,2,\ldots$, with $n$ fixed ($n$ indicates the order of the superstable orbit of period $2^n$, or that of the Misiurewicz point when $2^n$ bands are about to appear. Then we look at the scaling of these densities as the accumulation point of the superstable cycles is approached. We confirm the stationary character of the densities in the larger time scale $\tau$. In Section 4 the previous numerical results are reproduced via a rescaling scheme that uses as starting input the smooth, invariant distribution for the fully-developed single-band attractor at the end value of the map control parameter, known as the Ulam density \cite{Schuster1, Hilborn1, Beck1}. The rescaling procedure expresses the self-affinity that permeates through the properties of the logistic and other quadratic maps and reproduces sequentially the invariant densities at the band-splitting points (and also those for the supercycles) in a quantitative way on the time scale $\tau$. Next, in Section 5 we put together the properties of the two families of attractors studied, including their common accumulation point, into a Renormalization Group (RG) framework, such that the densities flow towards two trivial fixed points, period one for the periodic attractors and a single band for the chaotic attractors. The density for the accumulation point at the transition into or out of chaos corresponds to the nontrivial fixed point. Then we evaluate the (Shannon) entropies of the densities along the two families of attractors and observe that the entropy attains extreme values at the RG fixed points \cite{Robledo1, Luque1}, while the overall shape of the entropy and its derivative as a function of the map control parameter displays the characteristics of an equation of state and the response function of a statistical-mechanical system undergoing a second order phase transition. Finally, in Section 6 we summarize and discuss our results.\\


\section{\label{Sec:FokkerPlankLogMap}A Fokker-Planck equation for the logistic map}

The normal iterative procedure of generating trajectories $x_t, \  t = 0,1,2\ldots,$ from the logistic map

\begin{equation}
f_{\mu} (x)= 1 - \mu x^2, \ \ x \in [-1,1], \ \ \mu \in [0,2],
\label{Eq:LogisticMap_Def}
\end{equation}
starting from an initial condition $x_0$ at fixed value of the control parameter $\mu$, resembles the description of fluid motion in the Lagrangian frame of reference where $x_{t+1}=1-\mu x_t^2$ plays the role of a Langevin equation \cite{Chaikin1}. The change of perspective when looking at fluid motion through an Eulerian frame of reference involves the time evolution of the density of particles as represented by the Fokker-Planck equation \cite{Risken1}. The parallel description to study the dynamics contained in the logistic map centers on the transition probability of trajectories between positions reached at consecutive iteration times. This is   

\begin{equation}
\rho_t(x) = \int \mathrm{d}x' p(x,t|x',t-1)\rho_{t-1}(x'),
\label{Eq:Chapman-Kolmogorov_LogMap1}
\end{equation}
with $x = 1-\mu x'^2$. In our example, the conditional transition probability is given by the Dirac delta function

\begin{equation}
p(x,t|x',t-1) = \delta(x-1+\mu x'^2).
\label{Eq:TransitionProbability_LogMap}
\end{equation}
Therefore we have

\begin{equation}
\rho_t(x) = \int^1_{-1} \mathrm{d}x' \delta(x-1+\mu x'^2)\rho_{t-1}(x'),
\label{Eq:Chapman-Kolmogorov_LogMap2}
\end{equation}
which, after the change of variables $y = 1-\mu x'^2$ for $x'\in[0,1]$ and $z = 1-\mu x'^2$ for $x'\in[-1,0]$ becomes 

\begin{equation}
\rho_t(x) = \frac{1}{2\sqrt{\mu(1-x)}}\left[\rho_{t-1}\left(\xi_{\mu}(x)\right)+\rho_{t-1}\left(-\xi_{\mu}(x)\right)\right],
\label{Eq:Fokker-Planck_eqn_LogMap1}
\end{equation}
with $\xi_{\mu}(x) \equiv \sqrt{ (1-x)/\mu}$ and where $x\in [1-\mu,1]$ and $\rho_t=0$ for $x\in (-1,1-\mu)$. This is the Frobenius-Perron equation particularized to the logistic map when written for consecutive iteration times.\\

There is an important difference between the familiar linear Fokker-Planck equation in fluid motion or diffusion problems and the equation we obtained for the logistic map, Eq. (\ref{Eq:Fokker-Planck_eqn_LogMap1}), and this is that the functional inverse of the logistic map is not unique. Therefore, the ``backwards'' companion equation to Eq. (\ref{Eq:Fokker-Planck_eqn_LogMap1}), analogous to reverse time in the Fokker-Planck equation, is obtained by inserting in

\begin{equation}
\rho_{t-1} (x)= \int_{1-\mu}^1 \mathrm{d}x' p(x,t-1|x',t) \rho_t(x')
\label{Eq:Chapman-Kolmogorov_LogMap3}
\end{equation}
the backward propagation of the probability density, with $x = \pm \xi'_{\mu}$, $\xi'_{\mu} = \sqrt{ (1-x')/\mu}$. Explicitly, the conditional transition probability in Eq.(\ref{Eq:Chapman-Kolmogorov_LogMap1}) is a sum of two Dirac deltas 

\begin{equation}
p(x,t-1|x',t) = \frac{1}{2}\delta\left(x+\xi'_{\mu}\right)+\frac{1}{2}\delta\left(x-\xi'_{\mu}\right).
\label{Eq:TransitionProbability_LogMap3}
\end{equation}
Hence we have the expression

\begin{equation}
\rho_{t-1}(x) = \int^1_{1-\mu} \mathrm{d}x' \left[\frac{1}{2}\delta\left(x+\xi'_{\mu}\right)+\frac{1}{2}\delta\left(x-\xi'_{\mu}\right) \right]\rho_t(x'),
\label{Eq:Chapman-Kolmogorov_LogMap4}
\end{equation}
that yields

\begin{equation}
\rho_{t-1}(x) = \begin{cases}
			\ \ \ \mu x \rho_t(1-\mu x^2), \ \ x \in[0,1]\\
			-\mu x \rho_t(1-\mu x^2), \ \ x \in[-1,0],
		   \end{cases}
\end{equation}
which is already normalized.


\section{\label{Sec:EvoDensTraj} Evolution of densities of trajectories towards attractors}

\subsection{Period doubling cascade.}\label{Sec:DensityAtS}

The family of superstable attractors or supercycles \cite{Schuster1,Hilborn1} of the logistic map, and in general unimodal maps, has become a standard choice when describing dynamical properties along the period-doubling cascade. The rapid convergence of trajectories into these attractors (exponential of an exponential decay rate \cite{Robledo2}) was a convenient option in the early studies that revealed basic properties and defined key quantities, such as the so-called diameters \cite{Schuster1,Hilborn1}, and this in turn have stimulated many subsequent developments through their use. We select this family of attractors to determine densities of trajectories via the Frobenius-Perron method. The control parameter value for the supercycle of period $2^n$ is denoted $S_n, \ n = 1,2,3,\ldots$\\

\begin{figure}
\centering
\includegraphics[width=0.8\textwidth]{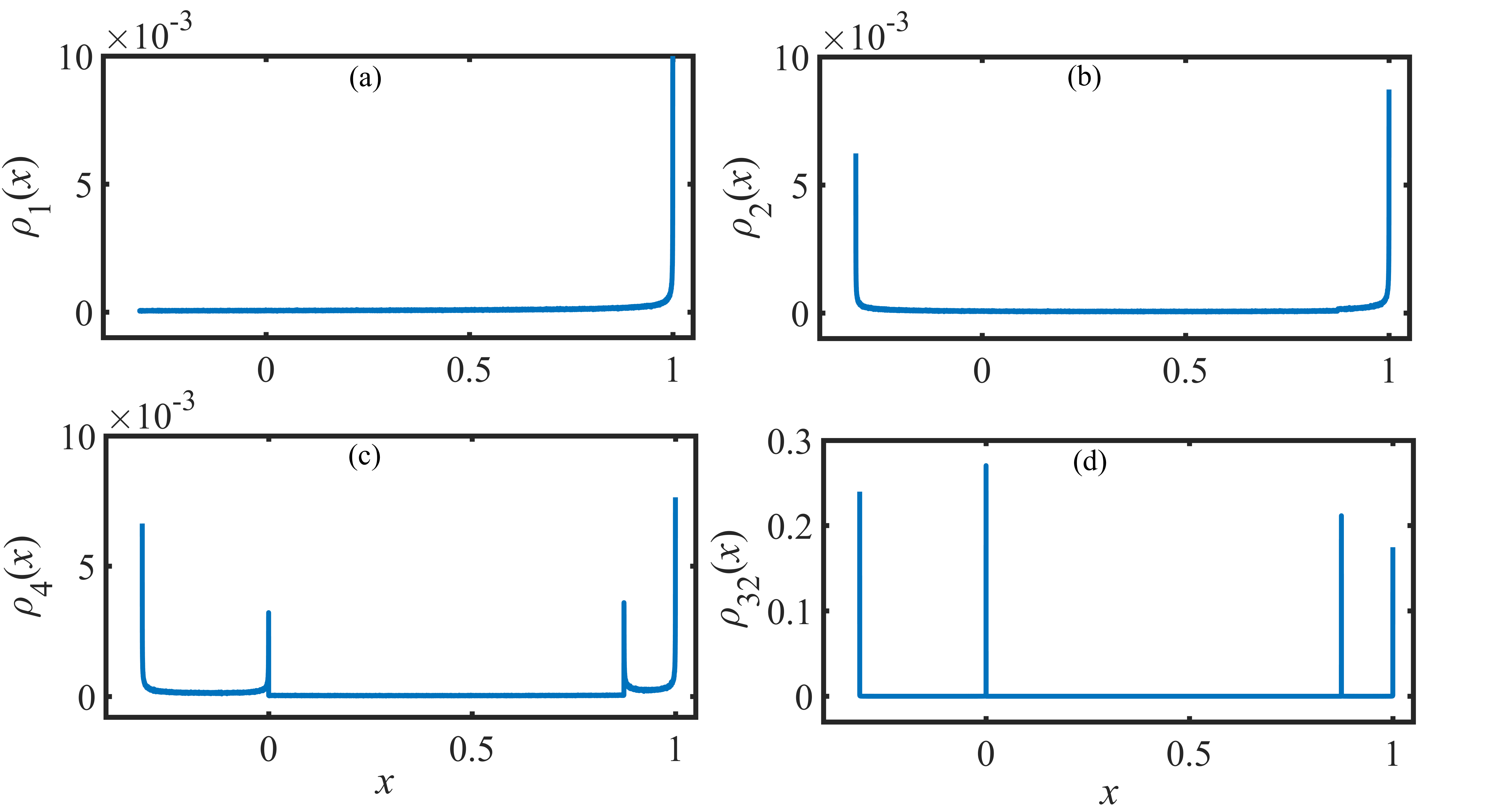}
\caption{Evolution of an initially uniform density of positions in the interval $[-1,1]$ when $\mu=S_2$. (a) First iteration, the density accumulates sharply around $x=1$. (b) Second iteration, another peak forms around $x \sim -0.3$. (c) Fourth iteration, there are four peaks centered on the final attractor points $x \sim-0.31, 0, 0.87, 1$. (d) Thirty two iterations, we observe already a good approximation of the final density consisting of four delta functions.} \label{Fig:Period4_4panels}
\end{figure}

In Fig. \ref{Fig:Period4_4panels} we show numerical results for the solutions of the Frobenius-Perron Eq. (\ref{Eq:Fokker-Planck_eqn_LogMap1}) when $\mu=S_2$ at early and moderately large iteration times for an initially uniform distribution of initial positions in the interval $[-1,1]$. As observed there, only one iteration is sufficient to wipe out uniformity and concentrate the trajectories near $x=1$. A second iteration divides the trajectories into two groups around two positions of the attractor with the formation of one central gap. Soon after the trajectories divide into four groups located close to the four attractor positions separated by two new gaps. The total of three gaps contain the three repellor positions present for period-four attractors. For all subsequent number of iterations the heights of the four peaks that express the populations of the four groups of trajectories alternate locations according to the fixed order of visits of attractor positions in unimodal map dynamics \cite{Schuster1}. For large iteration time $t$ the density is invariant when observed at multiples of $2^2 t$.  

\begin{figure}
\centering
\includegraphics[width=0.8\textwidth]{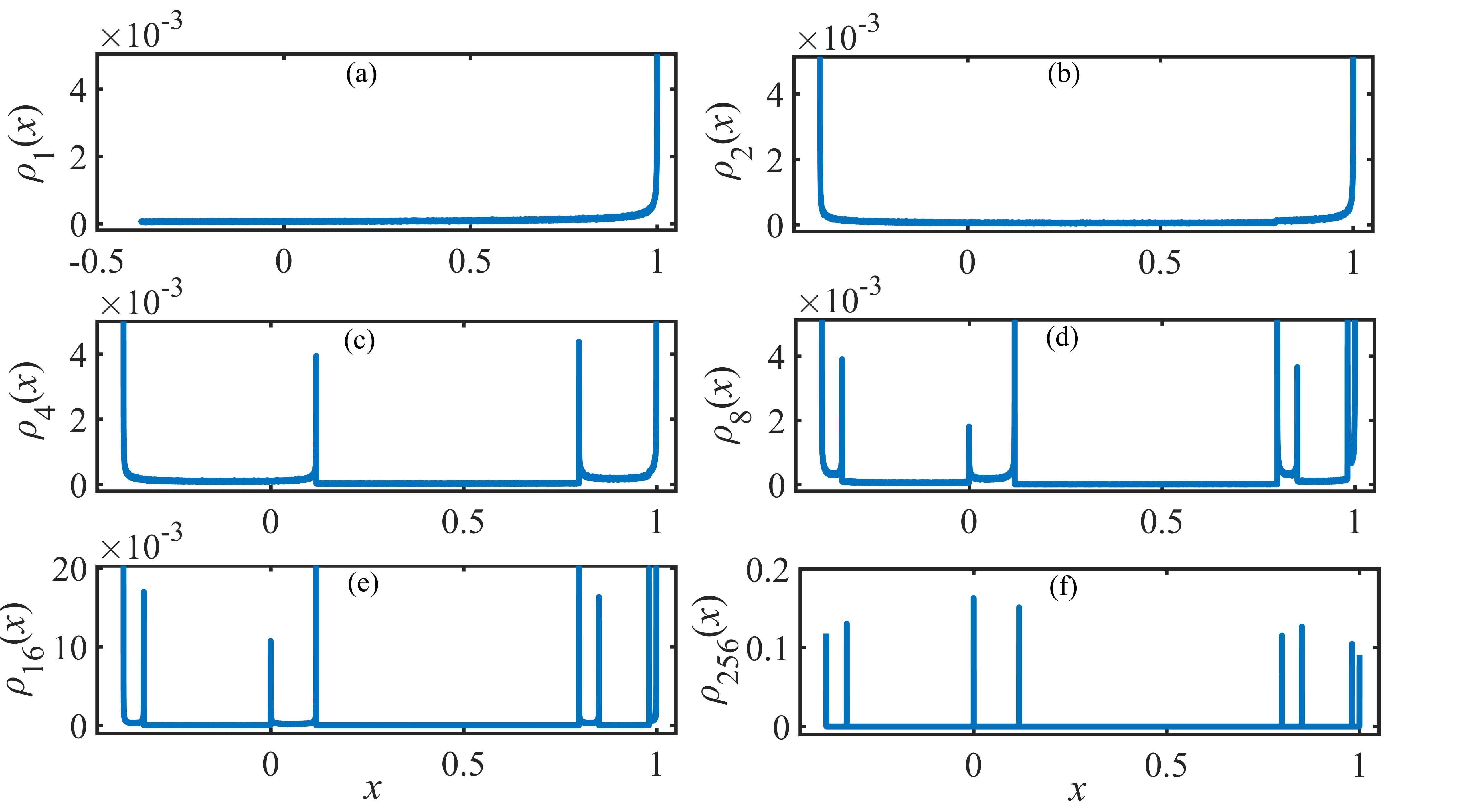}\caption{Evolution of an initially uniform density of positions in the interval $[-1,1]$ when $\mu=S_3$. (a) 1st iteration, the density rises steeply around $x=1$. (b) 2nd iteration, a second peak forms around $x \sim -0.3$. (c) 4th iteration, two more peaks develop between the former two. (d) 8th iteration, four additional peaks appear, there are eight peaks centered on the final attractor points  {$x \sim-0.38,  -0.33, 0, 0.12, 0.80, 0.85,0.98, 1$}. (e) 16th iteration, the gaps between the attractor positions develop further. (f) 256th iteration, we observe already a good approximation of the final density consisting of eight delta functions.}\label{Fig:Period8_6panels}
\end{figure}

Similarly, in Fig. \ref{Fig:Period8_6panels} we show numerical results for the solutions of Eq. (\ref{Eq:Fokker-Planck_eqn_LogMap1}) when $\mu=S_3$ at early and large iteration times for an initially uniform distribution of initial positions in the interval $[-1,1]$. As seen before, uniformity is sharply erased at the first iteration and trajectories cluster around $x=1$. Trajectories split at the second iteration into two groups centered around two attractor positions and create one central gap. Four groups of trajectories now around four attractor positions appear at four iterations while the central gap divides into three gaps. Lastly, eight groups of trajectories are formed on the eight attractor positions at iteration eight separated by seven gaps. The final sharp delta peaks that form the attractor separated by empty intervals are reached continuously for larger iteration times. As before, the heights of the eight peaks, that correspond to the populations of the eight groups of trajectories, alternate locations according to the fixed order of visits of attractor positions in unimodal map dynamics \cite{Schuster1}. For large iteration time $t$ the density appears invariant when observed at multiples of $2^3 t$.
 
For (all) larger periods the solutions of Eq. (\ref{Eq:Fokker-Planck_eqn_LogMap1}) at $\mu=S_n,\ n=4, \ldots$, show a parallel iteration time development. This is to recapitulate the formation of $2^k, k=0, 1, 2, 3, \ldots, n$, peaks and the, $2^k - 1, k=0, 1, 2, 3, \ldots, n-1$, gaps between them sequentially at times $t=2^k, k=0, 1, 2, 3, \ldots, n$. The large time $t\rightarrow\infty$ density is the sum of $2^n$ delta functions centered at the attractor positions $x_i, i=1,\ldots,n$. The density vanishes in the intervals between these positions that contain the $2^n - 1$ repellor positions. The amplitudes of the delta functions are in general dependent on the initial distribution of initial conditions (uniform in our case) for the ensemble of trajectories. These amplitudes retain forever an iteration time dependence as they are circulated amongst the $2^n$ attractor positions according to the prescribed order of visits in the dynamics of unimodal maps \cite{Schuster1, Hilborn1, Beck1}. The density is invariant when the alternative time scale $\tau=N 2^n, N=1,2,\ldots$, is adopted. The two time scales $t$ and $\tau$ diverge from each other exponentially as $n\rightarrow\infty$.

\subsection{Chaotic band-splitting cascade.}\label{Sec:DensityAtBS}

\begin{figure*}
\centering
\includegraphics[width=0.8\textwidth]{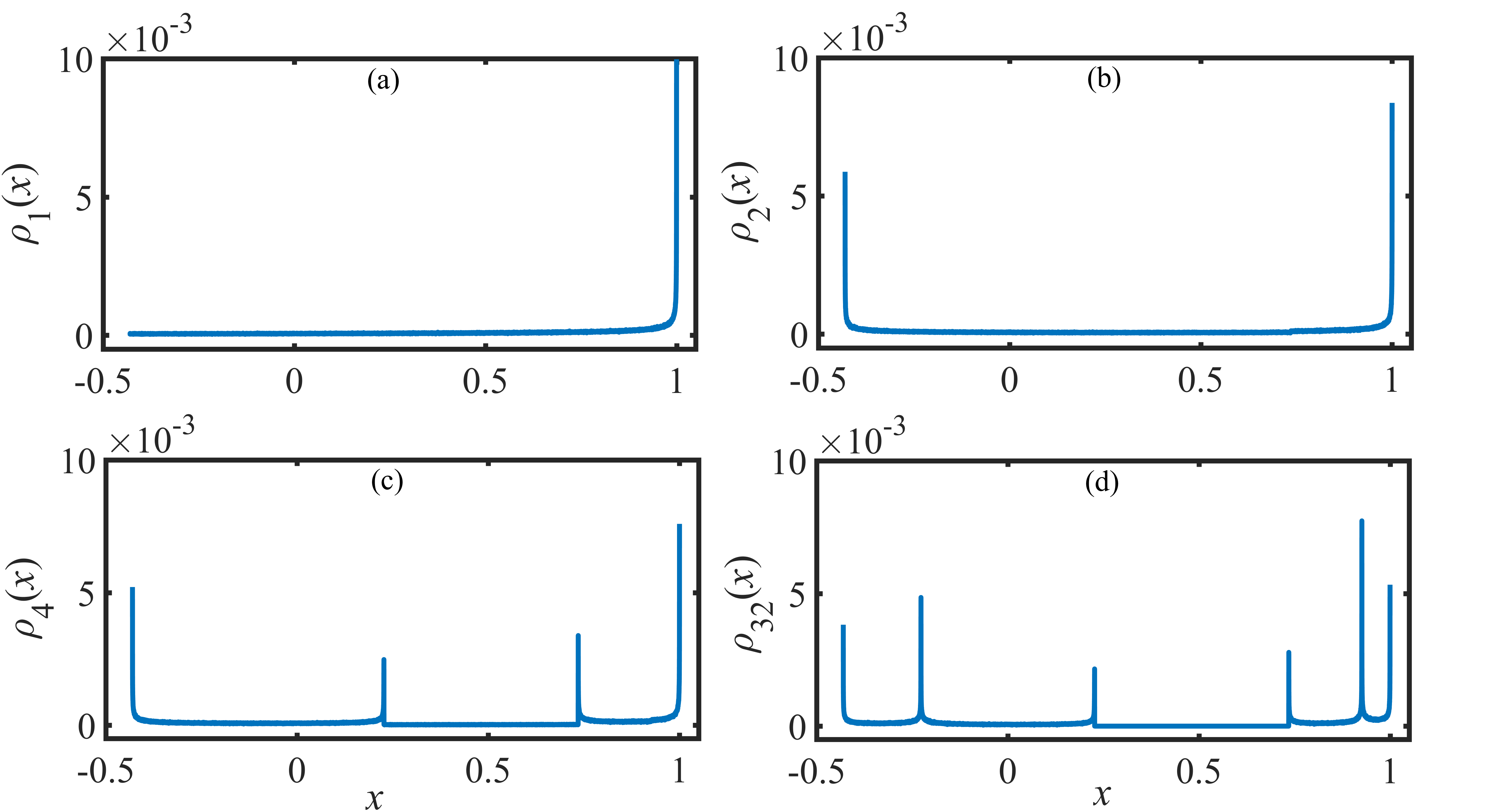}
\caption{Evolution of an initially uniform density of positions in the interval $[-1,1]$ when $\mu=M_2$. (a) First iteration, the density accumulates sharply around $x=1$. (b) Second iteration, another peak forms around $x \sim -0.4$. (c) Fourth iteration, there are four peaks delineating the edges of the two final bands separated by a gap. (d) Thirty two iterations, we observe already a good approximation of the final density that shows the twin $u$-shaped form at each of the two bands.} \label{Fig:M1_4panels}
\end{figure*}

\begin{figure*}
\centering
\includegraphics[width=0.82\textwidth]{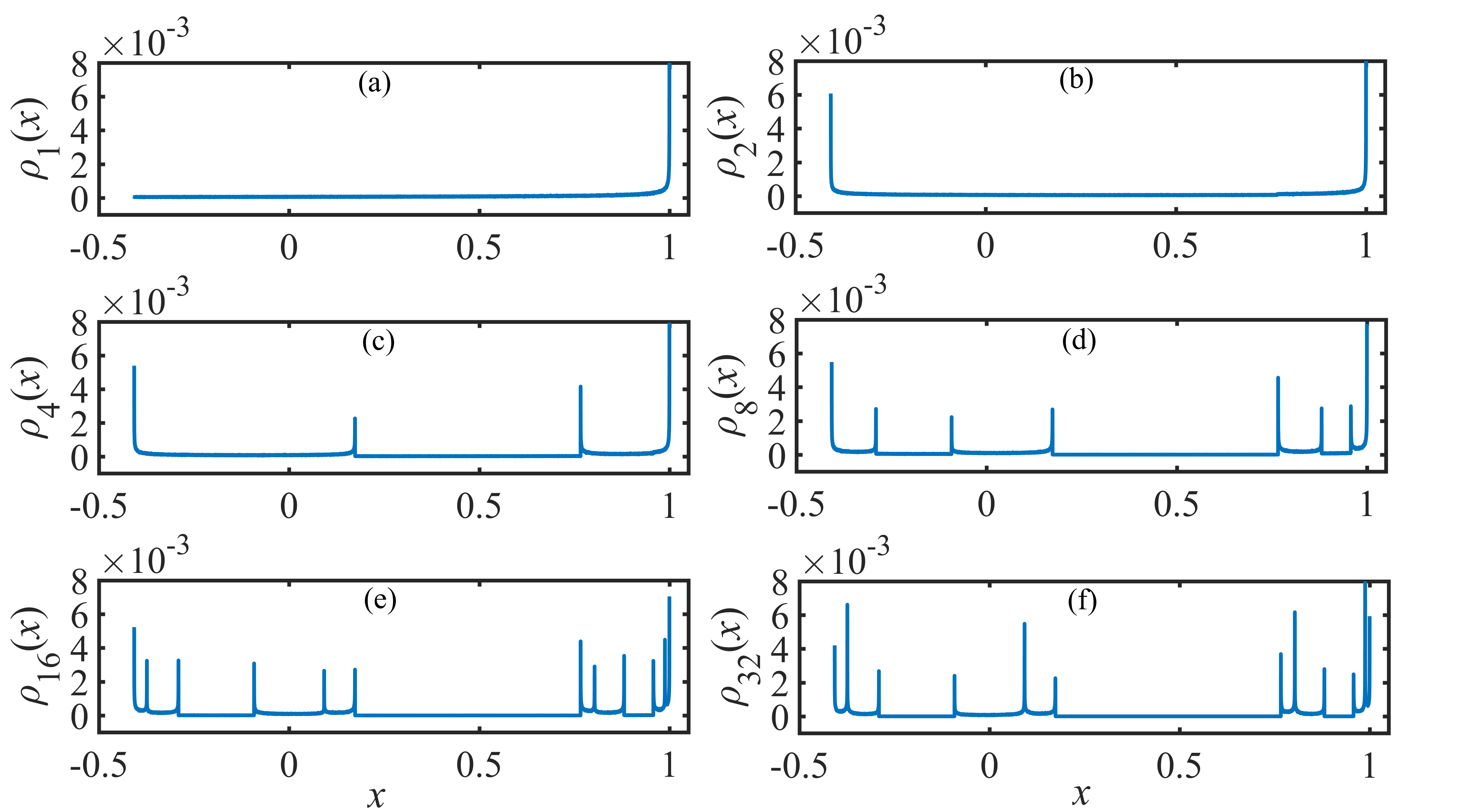}\caption{Evolution of an initially uniform density of positions in the interval $[-1,1]$ when $\mu=M_3$. (a) 1st iteration, the density rises steeply around $x=1$. (b) 2nd iteration, a second peak forms around $x \sim -0.4$. (c) 4th iteration, two more peaks develop between the former two forming two bands separated by a central gap. (d) 8th iteration, four additional peaks appear, leading to four bands separated by three gaps. (e) 16th iteration, the gaps between the attractor positions develop further while four new peaks give the densities the characteristic Misiurewicz-point repeated twin $u$-shaped form. (f) 32nd iteration, we observe already a good approximation of the final density.}\label{Fig:M2_6panels}
\end{figure*}

\noindent The family of chaotic-band attractors where bands are on the point to split, known also as Misiurewicz points \cite{Beck1} is a convenient option introduced here to determine densities of chaotic trajectories via the Frobenius-Perron method. The band-splitting sequence for $\mu>\mu_\infty$ is the chaotic equivalent to the period-doubling supercycles for $\mu<\mu_\infty$. For recent developments assisted through their use, see for example \cite{Alvaro1, Alvaro2}.  {The determination of the band widths at the $n$th Misiurewicz point is facilitated by the circumstance that the set of band edge points, that we denote by $\{e_{n,k}\} \in [1-\mu,1], \ k = 1,\ldots,3*2^{n-1}$, correspond to positions of the the trajectory initiated at the position of the map $x=0$, \textit{i.e.}, $e_{n,k} = f_{M_n}^k(0)$, where $M_n, \ n = 1,2,3,\ldots$ is the control parameter value for the $2^n$-band Misiurewicz point. Orbits initiated at the position $x=0$ for $\mu=M_n$ are \textit{eventually periodic} \cite{Romera1996}, which means they posses a transient called $preperiod$ after which they are periodic. In the former, $3*2^{n-1}$ is the sum of the preperiod $q=2^{n}$ and the period $p=2^{n-1}$ of each $M_n$. Therefore $e_{n,1} = 1,\ e_{n,2} = 1-M_n,\ e_{n,3} = 1-M_n e_{n,2}^2 = 1 - M_n(1-M_n)^2$, and so on. The positions of the points  $e_{n,k}$ correspond to the peaks of the density in the final panels in Figs. \ref{Fig:M1_4panels}\ and \ref{Fig:M2_6panels}. Following the same procedure for arbitrary values of $\mu$ defines the polynomials}
 
\begin{equation}
     P_{n}(\mu)=1-\mu P_{n-1}^2(\mu), \ P_0 = 1, \ \ \mu\in[0,2]
     \label{Eq:mu_Polynomials}
\end{equation}

\noindent  {of order $2^n-1$, that are sometimes referred to as shade curves or critical polynomials \cite{Romera1996}, that we will call here simply $\mu$-polynomials or $\mu$-curves. They conform the loci of all the band edges, and supercycle and periodic window's attracting positions (see Fig. \ref{Fig:BDcurves}), a fact that will be employed in the construction of the Renormalization-Group (RG) transformation.}
 
 Fig. \ref{Fig:M1_4panels} shows numerical results for the solutions of the Frobenius-Perron Eq. (\ref{Eq:Fokker-Planck_eqn_LogMap1}) when $\mu=M_2$ at early and larger iteration times for an initially uniform distribution of initial positions in the interval $[-1,1]$. Again, only one iteration is sufficient to wipe out uniformity and concentrate the trajectories near $x=1$. At a second iteration appears a second peak that gives the distribution a single $u$-shaped form. At the fourth iteration the trajectories are divided into two groups forming $u$-shaped densities separated by one central gap. A few more iterations lead to a density that approximates the final form for this chaotic band attractor. For large iteration time $t$ the density is invariant when observed at multiples of $2^2 t$. 

Similarly, in Fig. \ref{Fig:M2_6panels} we show numerical results for the solutions of Eq. (\ref{Eq:Fokker-Planck_eqn_LogMap1}) when $\mu=M_3$ at early and large iteration times for an initially uniform distribution of initial positions in the interval $[-1,1]$. Again, the same sequential pattern is observed, fast departure from uniformity with clustering first at $x=1$, then at the other edge of the main band, central gap formation, and splitting of the two bands. Next, the same events leading now to three gaps separating four bands, each exhibiting the characteristic double $u$-shaped density of the Misiurewicz points. For large iteration time $t$ the density is invariant when observed at multiples of $2^3 t$. 

\subsection{Density of iterates at the Feigenbaum point}\label{sSec:DensAtFeigPoint}

The evolution via sequential gap formation of uniformly distributed ensemble of trajectories towards supercycle and Misiurewicz point attractors display a `recapitulation' property \cite{Alvaro2, Alvaro3, Robledo3}, i.e. progression towards $2^n$-periodic or $2^n$-band chaotic attractors repeats successively that towards those attractors with $2^k , k = 0, 1, 2, ..., n-1$. As described above, we have seen that this property appears reflected in the time development of their densities. When $\mu=\mu_{\infty}$ recapitulation never ends and the attractor becomes a multifractal set while the density becomes an infinite set of delta functions placed at the attractor positions (see Fig. \ref{Fig:Zebra} below). At the accumulation point the difference between the two time scales $t$ and $\tau$ diverges and advancing by successive iterations does not reach the complete invariant density. A different option
is to place a uniform distribution on the multifractal attractor (say one initial condition on each point in Fig. \ref{Fig:Zebra}) and add a position at infinity, $x=x_{\infty}$, with the rule $f_{\mu_{\infty}} (x_{\infty})= 0$. This distribution remains invariant in the iteration time scale $t$.


\section{\label{Sec:DensityAtMS} Self-affinity of families of Invariant densities}

\subsection{Stationary density for $\mu=2$} \label{sSec:StationarySol_InvMeasure}

The invariant density of the Ulam map $f(x)=1-2x^2$ has been known for some time \cite{Schuster1, Hilborn1, Beck1}. For the logistic map in the fully chaotic regime $\mu=2$ the stationary solution $\rho_t(x) =\rho_{t+1}(x)$ is the $u$-shaped function

\begin{equation}
\rho (x) =\frac{1}{\pi\sqrt{1-x^2}},
\label{Eq:InvariantPDF_UlamMap}
\end{equation}  

\noindent as it satisfies

\begin{equation}
\rho(x) = \frac{1}{2\sqrt{\mu(1-x)}}\left[\rho \left(\sqrt{\frac{1-x}{\mu}}\right)+\rho\left(-\sqrt{\frac{1-x}{\mu}}\right) \right],
\label{Eq:Fokker-Planck_equilibrium}
\end{equation}

\noindent or

\begin{equation}
\rho(x) = \frac{\rho\left(\sqrt{\frac{1-x}{\mu}}\right)}{\sqrt{\mu(1-x)}},
\label{Eq:Fokker-Planck_equilibrium_UlamMap}
\end{equation} 

\noindent since Eq. (\ref{Eq:InvariantPDF_UlamMap}) is symmetric around $x=0$, $\rho(-x)=\rho(x)$.

\subsection{Scaling for Misiurewicz points}\label{Sec:ScalingInvDens_AtMisiurewicz}

Now, that we have determined numerically the invariant densities (in the large time scale $\tau=N2^n, N=1,2,\ldots$) at the Misiurewicz points we will reproduce them quantitatively by means of a scaling argument. Consider the affine transformation

\begin{equation}
y \equiv b x + c, \ 0<b<1, \ 1-\mu < c < 1 \label{Eq:AffineTransf_y}
\end{equation}

\noindent applied to the density $\rho_X(x)$, we have

\begin{eqnarray}
\rho_Y(y) &=& \int\mathrm{d}x \rho_X(x) \delta(y-b x-c)\\ \nonumber
	   &=& \frac{1}{b}\rho_X\left(\frac{y-c}{b}\right). \nonumber
	   \label{Eq:SelfAffineDensityTransformation}
\end{eqnarray}

\noindent When $\rho_X(x)$ is the Ulam invariant density (Eq. (\ref{Eq:InvariantPDF_UlamMap})) we obtain

\begin{equation}
\rho(x) = \frac{1}{\pi\sqrt{b^2-(x-c)^2}}.
\label{Eq:SelfAffine_Ulam}
\end{equation}

\noindent Considering that at a Misiurewicz point each chaotic band splits into two new bands, and that correspondingly the invariant density duplicates the number of $u$-shaped elements in it, then we assume that each $u$-shaped element at $M_n$ gets a proportion of $2^{-n}$ of the total measure at $\mu=2$. So, at the first Misiurewicz point the scaling ansatz gives

\begin{eqnarray}
\rho_{M_1}(x) = \begin{cases}
		\left(2\pi\sqrt{b_{1,1}^2-(x-c_{1,1})^2}\right)^{-1}, \ \ x\in[e_{1,3},e_{1,1}]\\
		\left(2\pi\sqrt{b_{1,2}^2-(x-c_{1,2})^2}\right)^{-1}, \ \ x\in[e_{1,2},e_{1,3}]
	   \end{cases}
	   \label{Eq:Density_M1}
\end{eqnarray}

\noindent  {where we introduce the notation $c_{n,i}, \ b_{n,i}$ with $n$ indicating the generation of the band-splitting cascade, and the second index denotes the $i$th $u$-shaped density element $i=\min(l,s)$ where $l,s=1,\ldots,3*2^{n-1}$ are the indices of the corresponding edges $e_{n,l},e_{n,s}$ of its support, that we denote by $U_{n,i}=[e_{n,l},e_{n,s}]$ following the same definition of its indices as above. In this way $c_{n,i}=(e_{n,l}+e_{n,s})/2$. The contraction parameters $b_{n,i}=|e_{n,l}-e_{n,s}|/2$ are central to the RG transformation that we elaborate in next section.}

\begin{figure}[htp]
\centering
\includegraphics[width=0.6\textwidth]{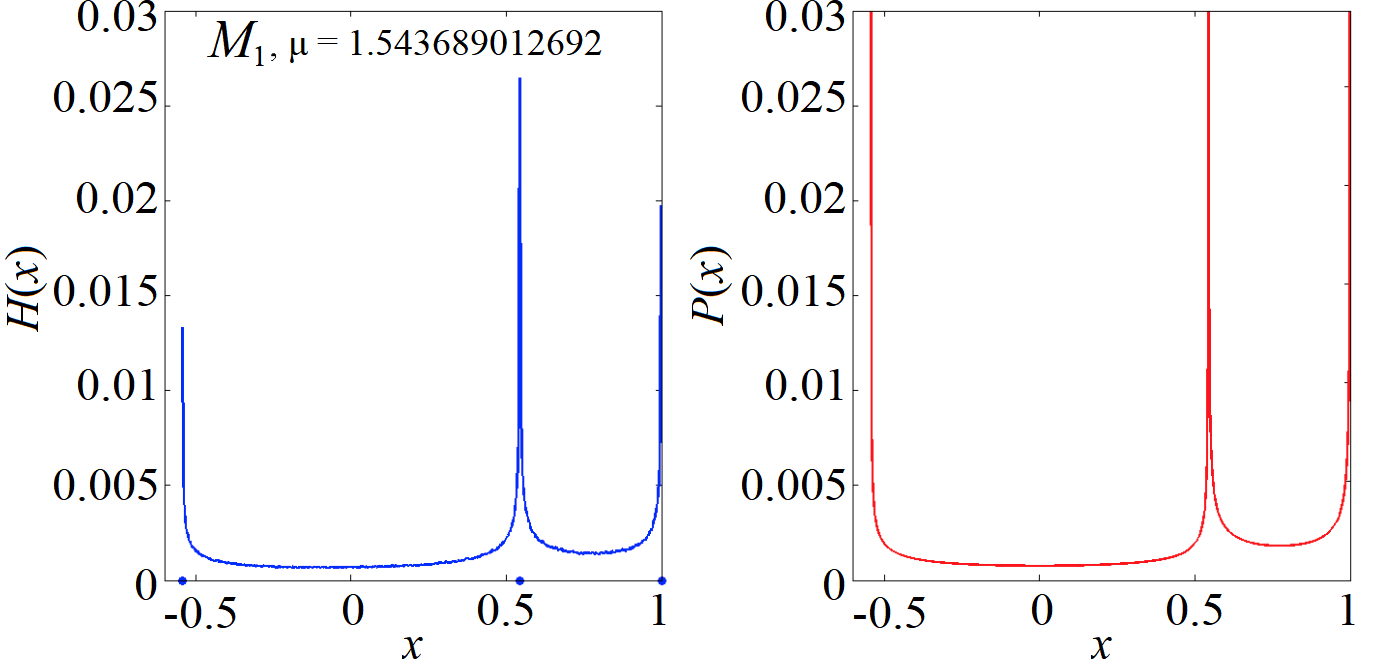}
\caption{Invariant density for the first Misiurewicz point $M_{1}$. Left panel, numerically determined from the Frobenius-Perron equation. Right panel, obtained from scaling and duplication of the Ulam density in Eq.(\ref{Eq:InvariantPDF_UlamMap}) according to Eq. (\ref{Eq:Density_M1}).}
\label{Fig:Self-AffineM1}
\end{figure}

\begin{figure}[htp]
\centering
\includegraphics[width=0.6\textwidth]{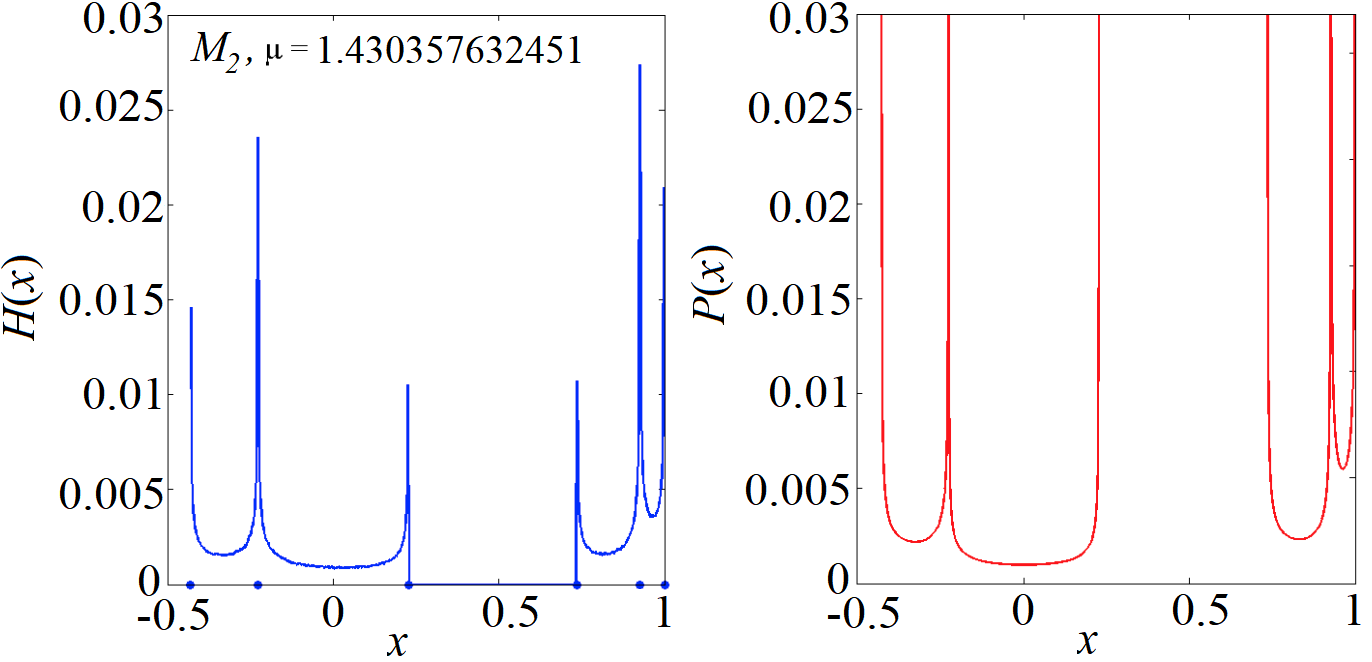}
\caption{Invariant density for the second Misiurewicz point $M_{2}$. Left panel, numerically determined from the Frobenius-Perron equation. Right panel, obtained from scaling and duplication of the density in Eq.(\ref{Eq:Density_M1}) according to the general expression in Eq. (\ref{Eq:AffineDensitiesEachM}).}
\label{Fig:Self-affineM2}
\end{figure}

\noindent  {Using the previous notation, we write the expression for the density at each Misiurewicz point $M_n$ as the Feigenbaum attractor is approached when $n\rightarrow\infty$. The $i$th density element of the measure at the $n$-th Misiurewicz point, with support $U_{n,i}$ has the form}

\begin{equation}
\rho_{n,i}(x) = \frac{1}{2^{n}\pi\sqrt{b_{n,i}^{2}-(x-c_{n,i})^{2}}}.
\label{Eq:AffineDensitiesEachM}
\end{equation}

\noindent Note that $b_{n,i}^2-(x-c_{n,i})^2\geq 0$ for $x\in U_{n,i}$. In Fig. \ref{Fig:Self-AffineM1} (Fig. \ref{Fig:Self-affineM2}) we show the agreement between the numerically determined and the scaled and duplicated invariant densities for the first (second) Misiurewicz point as given by Eq. (\ref{Eq:AffineDensitiesEachM}).


\section{\label{Sec:RG_transformation} Renormalization-Group transformation}

\noindent  {As it will be highlighted in the next section, the Feigenbaum point $\mu_\infty$ can be interpreted as the nontrivial fixed point of a \textit{discrete} RG transformation. The action of the transformation maps $\mu\neq\mu_\infty$ towards one of the trivial fixed points: either towards $\mu=0$ if $\mu\in \{ S_n\}$ or $\mu=2$ if $\mu=\{M_n \}$). For the sequence $\{ M_i \}$ of Misiurewicz points the direction of the RG flow is $M_n \rightarrow M_{n-1}$ and thus, it is given by the inverse of our self-affine transformation}

\begin{equation}
    Y(x)= \beta_{n,1} x + \gamma_{n,1} 
    \label{Eq:RGtransformation}
\end{equation}

\noindent  {with $Y(x)=y^{-1}(x)$, $\beta_{n,1}=1/b_{n,1}$ and $\gamma_{n,1}=-c_{n,1}/b_{n,1}$. For simplicity and clarity in the derivation, let us start by focusing on the intervals $U_{n,1}=[e_{n,q+1},e_{n,1}=1]$ (with $q=2^n$ the preperiod of $M_n$, see Sec. \ref{Sec:DensityAtBS}). The RG transformation maps the interval $U_{n,1}$ onto $U_{n-1,1}$ through Eq. (\ref{Eq:RGtransformation}). In order to realize this mapping, notice first that all the intervals $U_{n,1}$ share the same boundary at $x=1=e_{n,1}$. Mapping this edge  with Eq. (\ref{Eq:RGtransformation}) $e_{n+1,1}\rightarrow e_{n,1}$, gives the relation}

\begin{equation}
\gamma_{n,1}=1-\beta_{n,1}. 
\end{equation}

\begin{figure}[htp]
\centering
\includegraphics[width=0.75\textwidth]{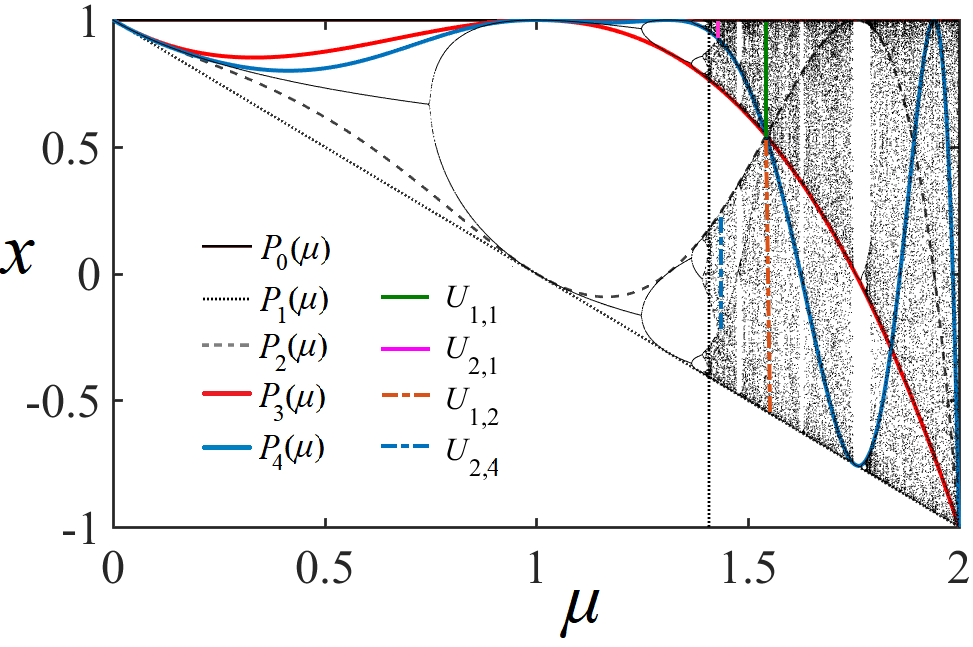}
\caption{  {Bifurcation diagram of the logistic map with superimposed $\mu$-polynomials $f^n_\mu (0)=P_n(\mu)=1-\mu P_{n-1}^2(\mu), \ P_0=1$ for $ n = 0,1,2,3,4$, outlining the bifurcation diagram. We indicate with vertical lines the intervals $U_{n,i}$ for $n={1,2}, \ i = {1,1,2,4}$, corresponging to the last steps on the discrete RG for $\mu \in \{ M_n\}$. The dotted vertical line indicates $\mu_\infty = 1.401155189092 $.}}
\label{Fig:BDcurves}
\end{figure}

\noindent  {For mapping the edges $e_{n,q+1}$ it is illustrative to write them first in terms of the $\mu$-polynomials defined by Eq. (\ref{Eq:mu_Polynomials}) (see also Fig. \ref{Fig:BDcurves}), evaluated at $\mu
_n=M_n$}

\begin{eqnarray}
    e_{n,q+1} = P_{q}(\mu_{n}) \label{Eq:edges_poly}, \nonumber
\end{eqnarray}

\noindent  {with $q= 2^{n}$. The mapping  $e_{n,q+1}\rightarrow e_{n,q^*+1}$ (with $q^*=2^{n-1}$) corresponds then to $P_{q}(\mu_{n})\rightarrow P_{q^*}(\mu_{n-1})$ or equivalently to the equation $Y(P_{q}(\mu_{n}))=P_{q^*}(\mu_{n-1})$, whose solution for $\beta_{n,1}$ is }

\begin{equation}
    \beta_{n,1} = \frac{1-P_{q^*}(\mu_{n-1})}{1-P_{q}(\mu_{n})}.
    \label{Eq:beta_n1}
\end{equation}

\noindent   {Notice that Eq. (\ref{Eq:beta_n1}) corresponds to the ratio of the lengths $|\cdot|$ of successive intervals $|U_{n-1,1}|/|U_{n,1}|$ and, we have obtained it through a RG argument. The asymptotic value $\lim_{n\rightarrow\infty}\beta_{n,1}=\beta_{\infty,1}$ is estimated with Eq. (\ref{Eq:beta_n1}) up to the 6th Misiurewicz point as $\beta_{6,1} = (1-P_{32}(M_{5}))/(1-P_{64}(M_{6})) \sim 6.26386840814597... $. This value has a discrepancy of only $0.011\% $  with respect to the square of a well known quantity: $\alpha^2 = 6.264547831212568$, $\alpha = -2.502907875095\ldots$ the universal Feigenbaum constant\cite{Feigenbaum1979} giving the local scaling around the maximum of all quadratic  unimodal maps at the accumulation point of the period-doubling scenario.}%

\begin{figure}[htp]
\centering
\includegraphics[width=0.6\textwidth]{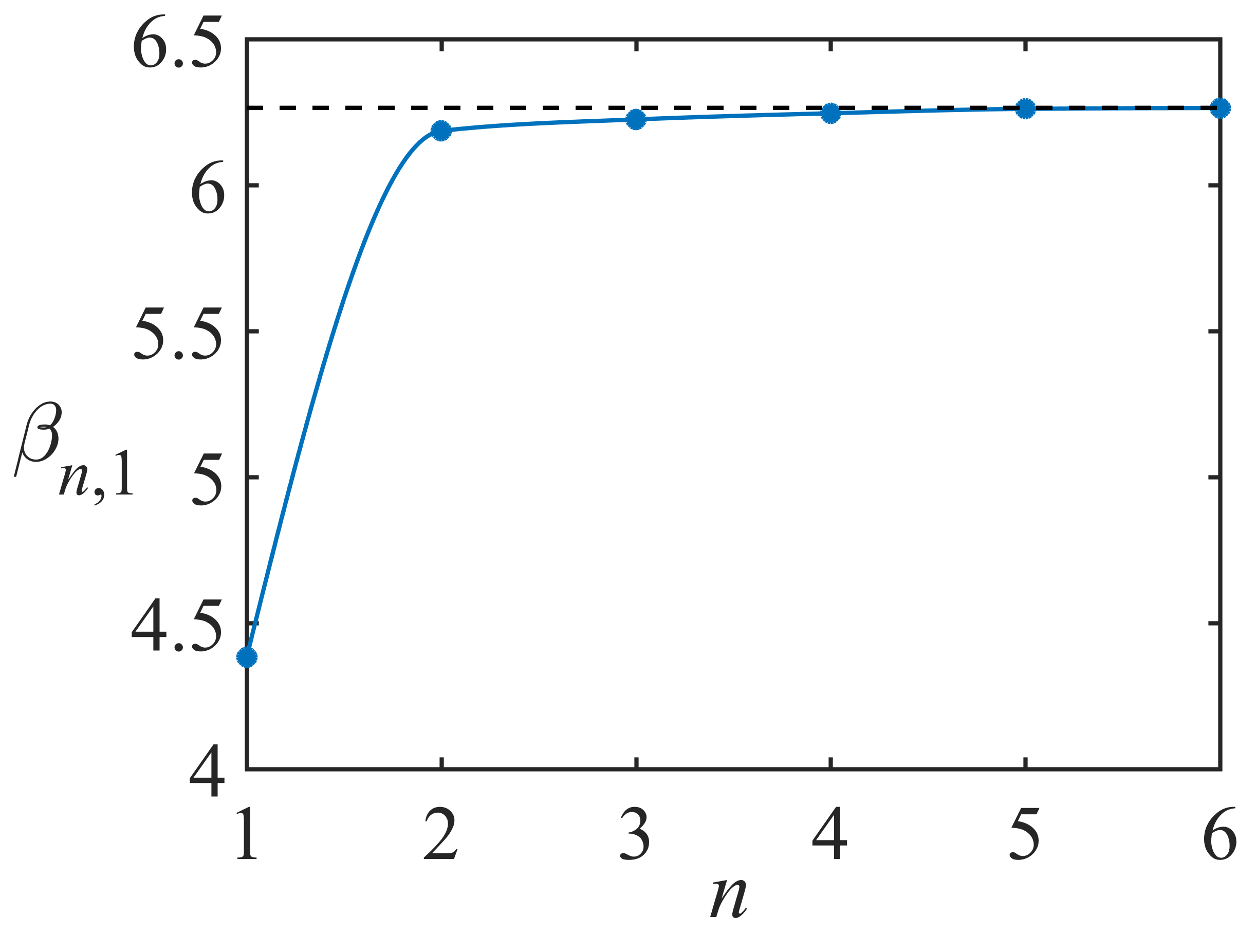}
\caption{ {Convergence of the expansion coefficients $\beta_{n,1}$ to the value $\beta_{\infty,1} = \alpha^2 =6.264547831217037\ldots$, indicated with a dashed horizontal line. The solid circles correspond to the first 6 control parameter values at Misiurewicz points. The line joining the points is only a guide to the eye.}}\label{Fig:bn}
\end{figure}

 {In Fig. (\ref{Fig:bn} we show the monotone convergence of the numerical estimate of $\beta_{n,1}$ to $\alpha^2$ as given by Eq. (\ref{Eq:beta_n1}).}

 {In a similar way, now we obtain the parameter $\beta_{n,q}$ for the symmetric intervals centered at $x=0$, $U_{n,q} = [e_{n,q*+q},e_{n,q}]$, with $e_{n,q}>0, \ e_{n,q*+q}<0$ and again $q*=2^{n-1}, \ q = 2^n$. This time $\gamma_{n,q}=0$, and by symmetry $e_{n,q} = -e_{n,q*+q}$ hence $|U_{n,q}| = 2 e_{n,q}$. The edge $e_{n,q}$ can be written in terms of the $\mu$-polynomials simply as $e_{n,q} = P_{q-1}(\mu_n)$ . With this, solving the RG equation $Y(P_{q-1}(\mu_n))=P_{q^*-1}(\mu_{n-1})$ for $\beta_{n,q}$ gives}

\begin{equation}
    \beta_{n,q} = \frac{P_{q*-1}(\mu_{n-1})}{P_{q-1}(\mu_n)}
    \label{Eq:beta_nq}
\end{equation}

 {By performing the corresponding numerical evaluations of Eq. (\ref{Eq:beta_nq}) for $\mu_n = \{ M_1,M_2,...,M_6\}$ we get $\beta_{6,64} = -2.50257123683208 $, this means a discrepancy of $0.013\%$ with respect to $\alpha$. Fig. \ref{Fig:bn} shows the values of the transformation coefficients $\beta_{n,1}$ for the first few $M_n$ as they converge to the limiting value $\lim_{n\rightarrow\infty} \beta_{n,1} = \alpha^2$. Our arguments for the derivation of the RG transformation and the numerical evidence for the convergence of $\beta_{n,1}\rightarrow\alpha^2$ and $\beta_{n,q}\rightarrow\alpha$ as $n$ is increased, enables us to conclude with confidence, that in the limit $n\rightarrow\infty$}

\begin{eqnarray}
\beta_{\infty,q} &=& \alpha \\
\beta_{\infty,1} &=& \alpha^2
\end{eqnarray}

 {The asymptotic values given above are naturally expected from the local scaling at $\mu_\infty$ around $x=0$ and $x=1$ for $\beta_{\infty,q}$ and $\beta_{\infty,1}$ , respectively, as given by the (recirpocal of) Feigenbaum's universal trajectory scaling function\cite{Schuster1,Feigenbaum1980} $1/\sigma(x=1)=\alpha^2$, indicating the most crowded region of the multifractal attractor, and $1/\sigma(x=0)=\alpha$ being the sparsest. This direct connection with the function $1/\sigma(x)$ provides an even more interesting interpretation to the asymptotic values $\beta_{\infty,i}$ and invites its reformulation as the function $\beta(x)=1/\sigma(x)$ in terms of the continuous variable $x=i/p$ with $p=2^{n-1}$ the period of $M_n$, just in the way it is done for $1/\sigma(x)$, thus providing a new way to obtain $1/\sigma(x)$ approaching form $\mu>\mu_\infty$.}
 {The designed RG transformation works also for $\mu<\mu_{\infty}$, \textit{i.e.}, at the sequence of supercycle attractors $\mu \in \{ S_n\}$.}


\section{\label{Sec:RGEntCrit}Renormalization Group, entropy, and criticality}

Instead of following the customary analytical format for the functional composition Renormalization Group (RG) procedure applied to the period doubling cascade \cite{Schuster1} we follow a graphical representation that facilitates its extension to the collection of the invariant densities we have already determined. Then we look at the entropy associated with them, and after this we remark on a statistical-mechanical critical point perspective of the Feigembaum accumulation point and its neighborhood.

\begin{figure}[htp]
\centering
\includegraphics[width=0.45\textwidth]{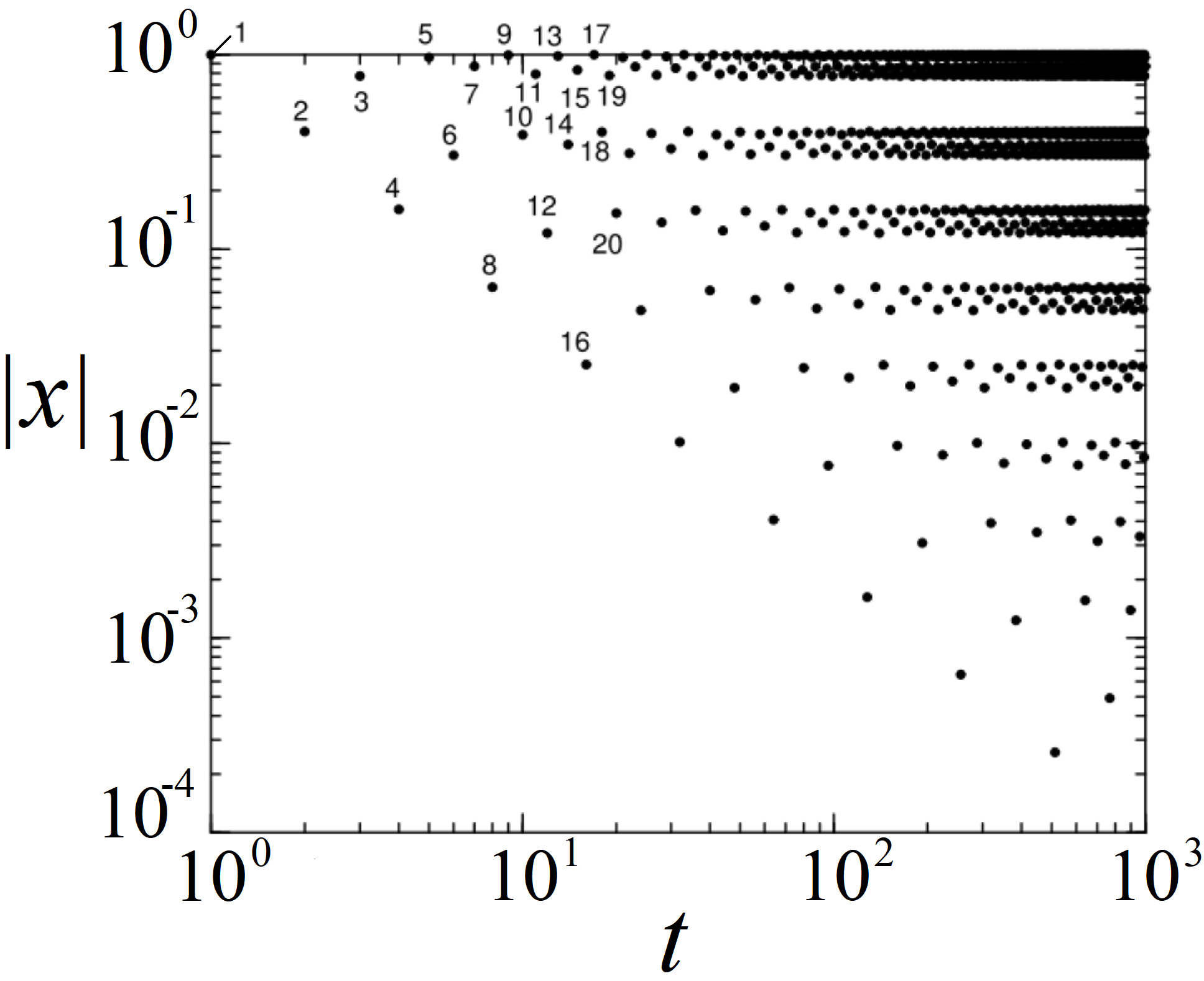}
\caption{Trajectory with initial condition $x=0$ at the Feigenbaum attractor $\mu_{\infty}$ in absolute values and logarithmic scales. From this visualization of this orbit it is straightforward to see how the absolute values of the positions group: All the odd iterates group to form the top band around $x=1$ and the rest follow different groupings into subsequent bands. We can appreciate also how subsequences with power law scaling are formed. Of special interest is the time subsequence corresponding to the powers $\{2^k\}$. See text.}\label{Fig:Zebra}
\end{figure}

In Fig. \ref{Fig:Zebra} we show the absolute values of positions in logarithmic scales of the first 1000 iterations for the trajectory initiated at $x_0$ when $\mu=\mu_{\infty}$. We observe in this figure that the positions appear arranged into horizontal bands separated by gaps, all bands of equal widths and all gaps of equal widths (seen more clearly defined for large $t$). The top band of positions contain $1/2$ of the positions, the next band $1/4$ of the positions, and so on, the $n$-th band $2^{-n}$ positions. The RG transformation is: i) Eliminate the top band, all positions with odd iteration times. This is half of the multifractal attractor. ii) Then shift the remaining positions to the left a distance $\ln 2$ and up a distance $\ln \alpha$. The result is that one recovers the same figure when  $\mu=\mu_{\infty}$. Repeat the operation any number of times. This is the nontrivial fixed point. Elimination of the  top band is the same as functional composition, so that the operations above correspond to the original RG. If $\mu$ is less than $\mu_{\infty}$, say we are at a supercycle, then the repeated RG operations lead to period one, one of the trivial fixed points. If $\mu$ is greater than $\mu_{\infty}$, say we are at a Misiurewicz point, then the repeated RG operations lead to one chaotic band, the other trivial fixed point.The RG in our plan is to do the same geometrical operations with the invariant densities obtained from the Frobenius-Perron, or, equivalently,  from the self-affine property. The (only) relevant variable (in RG language) is the difference $\Delta \mu \equiv \mu-\mu_{\infty}$, and it is similar to the temperature distance to the critical point in thermal systems. The RG relevant variables need to be set to zero in order to reach the nontrivial RG fixed point, the accumulation point at $\mu_{\infty}$. Otherwise the transformation flows towards the trivial RG fixed points, in our case period one or one chaotic band.

\subsection{A Renormalization Group scheme for invariant densities}\label{Sec:RGandFPsub}

The renormalization scheme operating on the invariant distributions at Misiurewicz points consists on folding each pair of adjacent $u$-shaped elements into one $u$-shaped unit followed by elimination of the gaps between them. Fig. \ref{Fig:Self-affineM2} for $M_2$ becomes Fig. \ref{Fig:Self-AffineM1} for $M_1$ under this transformation. The RG transformation works inversely with respect to the affine transformation in the previous section. The RG transformation for the (multi-delta function) invariant distributions at the supercycle attractors consists of merging pairs of their latest generation of delta functions into single ones, therefore eliminating the gaps between them and resulting into the invariant distribution of the previous supercycle. Fig. \ref{Fig:Period8_6panels}(f) for $S_3$ becomes Fig. \ref{Fig:Period4_4panels}(d) for $S_2$ under this transformation. Recall that the distributions for Misiurevicz Points $M_k$ and supercycle points $S_k$ are invariant in the time scale $\tau=N2^k, N=1,2,\ldots$, $k$ fixed, but show a cyclical pattern along iteration times $t$, one cycle covered through $t=N2^k, N2^{k}+1, N2^{k}+2, \dots, N2^{k}+2^k$, $N$ fixed. This, of course, after the transient behavior is over and only the asymptotic solution of the FP equation is observed.

\begin{figure}[htp]
\centering
\includegraphics[width=0.4\textwidth]{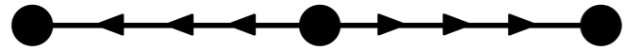}
\caption{Flow diagram of the renormalization group (RG) transformation. The RG transformation applied to any Misiurewicz point $M_k$ densities ($\Delta\mu>0$) leads to the trivial fixed point that represents the single band Ulam distribution, denoted by the right full circle. On the other hand, the RG transformation applied to any supercycle $S_k$ density ($\Delta\mu<0$) ends up at the trivial fixed point represented by the single delta function for period one, denoted by the left full circle. The non-trivial fixed point corresponds to the density made of an infinite set of delta functions located each at the positions shown in Fig. \ref{Fig:Zebra}, denoted the central full circle ($\Delta\mu=0$). See text.} \label{Fig:RGdiagram}
\end{figure} 

\subsection{Entropy and phase transition}

\begin{figure}[htp]
\centering
\includegraphics[width=0.6\textwidth]{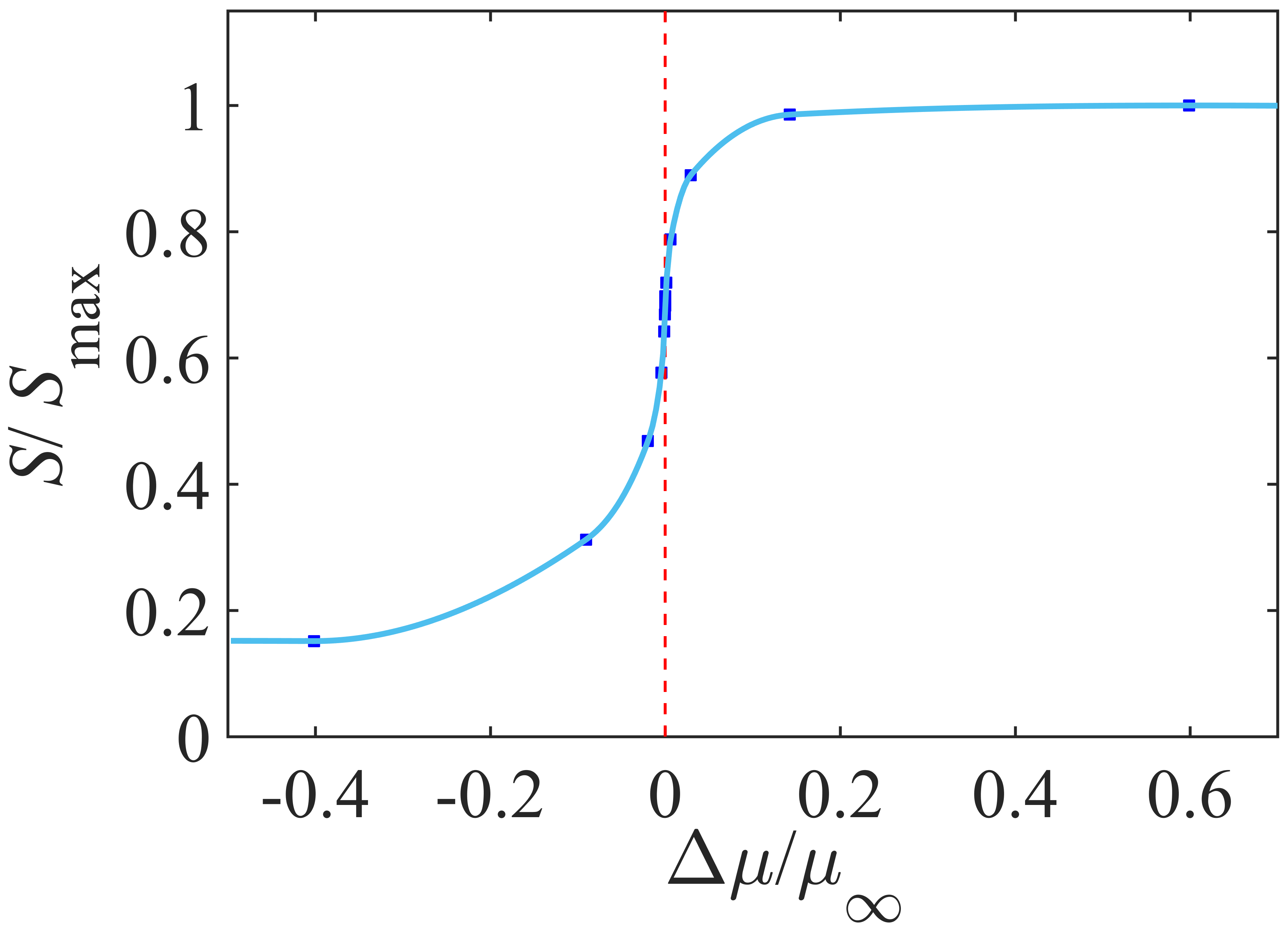}
\caption{Entropy vs control parameter value. The red, dotted vertical line represents the control parameter value $\mu_{\infty}$ of the transition to chaos, whereas each dot corresponds to a member of the sequences of either superstable orbits (left to the dotted line) or the band-splitting cascade (to the right of the dotted line).}\label{Fig:SvsMu}
\end{figure}

The Shannon entropy

 \begin{equation}
S_{\mu_{k}} = -\int^1_{-1} dx \rho_{\mu_{k}} \ln \rho_{\mu_{k}}(x),
\end{equation}

\noindent associated with the invariant densities $\rho_{\mu_{k}}$ at the families of supercycle attractors $\mu_{k}=S_{k}$ and Misiurewicz points $\mu_{k}= M_k$ we have determined can be readily obtained. These are shown in Fig. \ref{Fig:SvsMu} as a function of the control parameter distance to the accumulation point $\mu = \mu_\infty$. There we see behavior reminiscent of the entropy below, at, and above the critical temperature of a thermal system presenting ordered and disordered phases separated by a phase transition. However here we are following the entropy of ensembles of positions between two distinctive behaviors: The regular motion associated with the period-doubling cascade and irreguSolution of the RG eq gives:

\begin{equation}
    \beta_{n,1} = \frac{1-P_{q^*}(\mu_{n-1})}{1-P_{q}(\mu_{n})}.
    \label{Eq:beta_n1}
\end{equation}
by the former arguments and numerical evidence we conclude
\begin{equation}
\lim_{n\rightarrow\infty}\beta_{n,1} = \alpha^2
\end{equation}lar motion associated with the chaotic band-splitting cascade. The entropy presents a sudden increase at the transition from periodic motion to chaos. The logistic map on its route to chaos by either period-doubling or band splitting out of chaos can be viewed as a macroscopic system approaching a phase transition by a succession of equilibrium states. \\

If we recall that the RG trivial fixed points are the period-one supercycle and the single-band chaotic Ulam attractor and that the RG nontrivial fixed point is their common accumulation point at $\mu_{\infty}$ we notice that these points are also entropy extrema. The entropy at $\mu_{\infty}$ is maximum with respect to all periodic attractors and a minimum with respect to all chaotic attractors. This evidence reaffirms the claim advanced in past works \cite{Robledo13}, that the fixed points of the RG approach are always related to entropy extrema, with the all-important nontrivial fixed point as saddle point.  {See Fig. (\ref{Fig:RGdiagram})}. 

\begin{figure}[htp]
\centering
\includegraphics[width=0.6\textwidth]{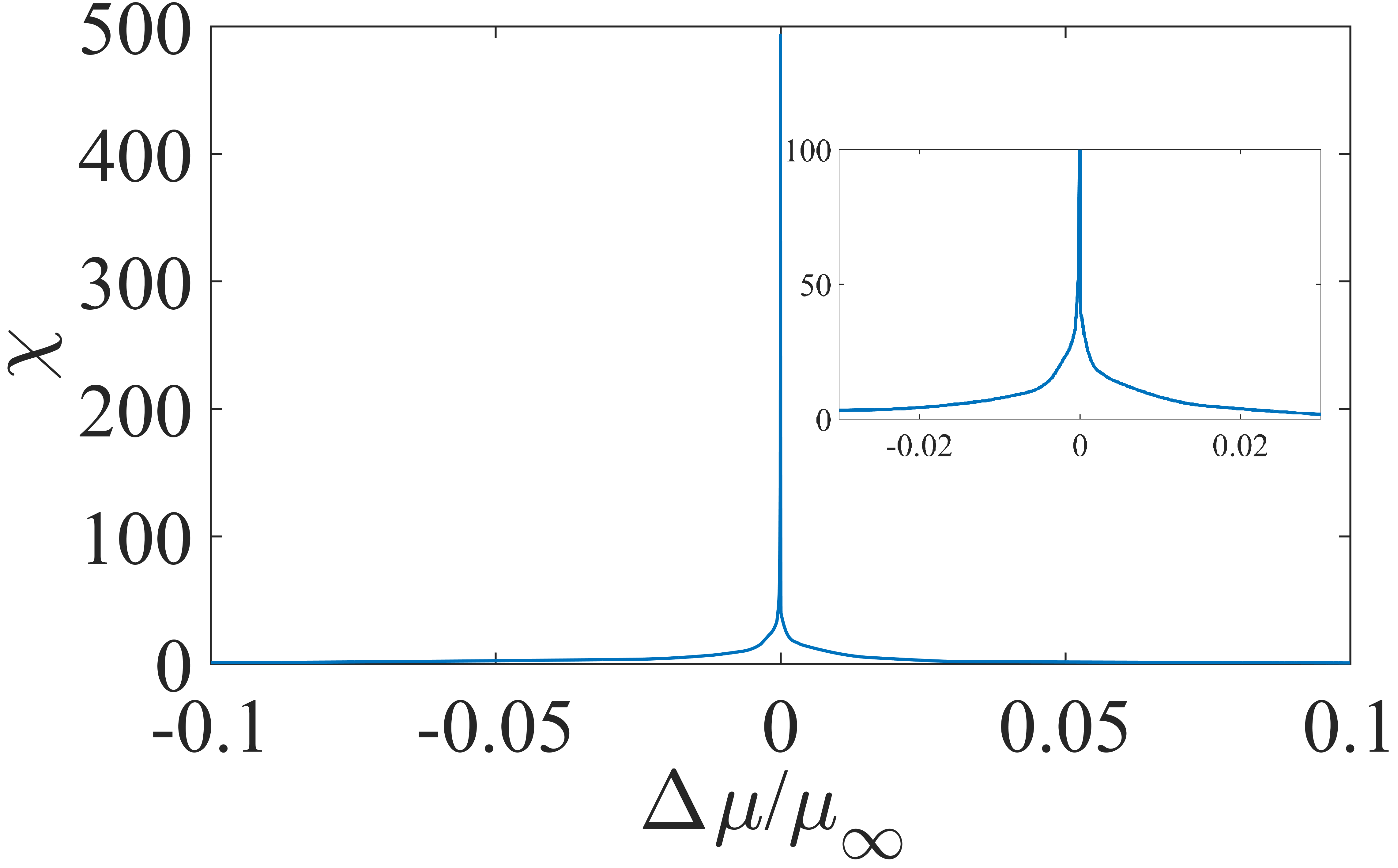}
\caption{Susceptibility $\chi$ calculated according to $\chi = \mu^{-1} \left| \partial \mu / \partial S \right|$ vs reduced distance in control parameter $\Delta\mu/\mu_{\infty}$. The characteristic divergence at a critical point is clearly appreciated at the transition to chaos $\Delta\mu = 0$.}\label{Fig:Susceptibility}
\end{figure}

To visualize further the parallelism with a thermal system we calculated from the data in Fig. \ref{Fig:SvsMu} the quantity that would correspond to a specific heat or susceptibility,

\begin{equation}
\chi =  \mu^{-1} \left|\frac{\partial S}{\partial \mu}\right|.\label{Eq:Susceptibility}
\end{equation}
\\
In Fig. \ref{Fig:Susceptibility} we can appreciate how $\chi$, as defined above, diverges at the onset of chaos.


\section{\label{Sec:SummDisc}Summary and discussion}

We examined the properties of the logistic map through the use of the Frobenius-Perron (FP) equation. We solved this equation numerically for both the first dozen supercycles along the period-doubling cascade and the first dozen Misiurewicz points along the chaotic band-splitting cascade of attractors. In both cases we observed a fast convergence to the final dynamical cyclical repetition. As a starting point we chose each time a uniform distribution of initial conditions in the interval of definition of the map. As expected when working with supercycles we observed very fast approach of the FP densities to their limiting form for trajectories inside the attractors. But this was also the case for the Misiurewicz points. We would expect to have observed a distinctively slower approach for the case of the pitchfork bifurcation points, while from our current experience we cannot indicate  what family of chaotic attractors, if any, would exhibit slow approach.

Only the Ulam distribution for the chaotic single band when $\mu=2$ and the single delta distribution for period one attractors are invariant in the consecutive iteration time scale $t$. All the distributions for Misiurewicz points $M_n$ and supercycles $S_n$  become invariant on the consecutive cycle time scale $\tau=N2^n, N=1,2,\ldots$, $n$ fixed. The two scales diverge from each other when the accumulation point of both families of attractors is approached, and so, the observation of an invariant density becomes increasingly unreachable. But this is not necessarily so if the set of initial conditions for the ensemble of trajectories is suitably chosen. A uniform distribution of initial conditions placed only in the attractor positions leads to an invariant density in the time scale $t$.

The invariant distributions obtained numerically from the FP equation were show to be quantitatively reproduced via a self-affine transformation with mirror duplication of either the Ulam distribution for all Misiurewicz points or of a single delta function for all supercycles. In addition to this a Renormalization Group (RG) transformation on the invariant densities defined as the reverse self-affine transformation with merging of mirror elements was introduced such that the RG flows towards two trivial fixed points that represent the distributions of a single fully-chaotic band and of a single periodic point. The nontrivial fixed point distribution is an infinite set of delta functions with multifractal support, the Feigenbaum attractor. The only relevant variable is the control parameter distance to the period-doubling accumulation point, $\Delta \mu \equiv \mu-\mu_{\infty}$.

The (Shannon) entropies $S$ for the distributions of the supercycles and the Misiurewicz points were determined and examined as a function of $\Delta \mu$. The outcome bears a strong resemblance with a critical isotherm in a typical thermal system undergoing a continuous phase transition. To affirm further this similarity the quantity that would correspond to a response function, $\chi = \mu^{-1} \left| \partial \mu / \partial S \right|$, was also calculated and was confirmed to display its characteristic divergence at a critical point. The families of attractors that form the bifurcation diagram of the quadratic maps show basically two behaviors, periodic and chaotic, appearing along two main cascades with increasing period or number of chaotic bands. They share the same accumulation point. These types of attractors represent the two `phases'  separated by a `critical' point. It is well-known that this feature is repeated an infinite number of times within the `periodic windows' in the fractal bifurcation diagram.  
\\
\\
AR acknowledges support from DGAPA-UNAM-IN106120 and Ciencia-de-Frontera-CONACyT-39572 (Mexican Agencies). 

\bibliography{ARFPFP}
\bibliographystyle{unsrt}

\end{document}